\documentclass[preprint, showpacs,preprintnumbers,amsmath,amssymb,nofootinbib]{revtex4}
\usepackage{amssymb}
\usepackage{amsfonts}

 \usepackage{epsf}
 \usepackage{graphicx}
\usepackage{float}
\begin{document}
\newcommand{\be}[1]{\begin{equation}\label{#1}}
 \newcommand{\ee}{\end{equation}}
 \newcommand{\bea}{\begin{eqnarray}}
 \newcommand{\eea}{\end{eqnarray}}
 \newcommand{\bed}{\begin{displaymath}}
 \newcommand{\eed}{\end{displaymath}}
 \def\disp{\displaystyle}

 \def\gsim{ \lower .75ex \hbox{$\sim$} \llap{\raise .27ex \hbox{$>$}} }
 \def\lsim{ \lower .75ex \hbox{$\sim$} \llap{\raise .27ex \hbox{$<$}} }

\title{A possible resolution of  tension between {\it Planck} and Type Ia supernova observations}

\author{Zhengxiang Li$^{1}$, Puxun Wu$^{3}$, Hongwei Yu$^{2, 3,} \footnote{Corresponding author: hwyu@hunnu.edu.cn}$, and Zong-Hong Zhu$^{1}$}

\address{$^1$Department of Astronomy, Beijing Normal University,
Beijing 100875, China
\\$^2$Department of Physics and Key Laboratory
of Low Dimensional Quantum Structures and Quantum Control of
Ministry of Education,
\\Hunan Normal University, Changsha, Hunan 410081, China
\\$^3$Center of Nonlinear Science and Department of Physics, Ningbo
University,  Ningbo, Zhejiang 315211, China}

\begin{abstract}
There is an apparent tension between cosmological parameters
obtained from {\it Planck} cosmic microwave background radiation
observations and that derived from the observed magnitude-redshift
relation for the type Ia supernova (SNe Ia). Here, we show that the
tension can be alleviated, if we first calibrate, with the help of
the distance-duality relation, the light-curve fitting parameters in
the distance estimation in SNe Ia observations with the angular
diameter distance data of the galaxy clusters and then re-estimate
the distances for the SNe Ia with the corrected fitting parameters.
This was used to explore their cosmological implications in the
context of the spatially flat cosmology. We find  a higher value for
the matter density parameter, $\Omega_m$, as compared to that from
the original SNLS3, which is in agreement with {\it Planck}
observations at 68.3\% confidence. Therefore, the tension between
{\it Planck} measurements and SNe Ia observations regarding
$\Omega_m$ can be effectively alleviated without invoking new
physics or resorting to extensions for the standard concordance
model. Moreover, with the absolute magnitude of a fiducial SNe Ia,
$M$, determined first, we obtained a constraint on the Hubble
constant with SNLS3 alone, which is also consistent with {\it
Planck}.

\end{abstract}


 \maketitle
 \renewcommand{\baselinestretch}{1.5}

\section{INTRODUCTION}

The cosmic microwave background radiation (CMBR) measurements play a
crucial and irreplaceable role in establishing the favored
cosmological model, that is, a flat cosmological constant-dominated,
cold dark matter model ($\Lambda$CDM), and constraining the
cosmological parameters. It is important, however, to bear in mind
that CMBR observations predominantly probe the early universe at
high redshift ($z\sim1100$). As a result, a projection within a
given cosmological model is needed when we interpret these
observations in terms of the standard cosmological parameters
defined at $z=0$, for instance, the Hubble constant, $H_0$, and the
matter density parameter, $\Omega_m$, which provide basic
information and are key parameters of the universe. Recently, one of
the most exciting events is the release of scientific findings based
on data from the first 15.5 months of {\it Planck}
operations~\cite{Planck1}. Because of the high precision, the new
{\it Planck} data could constrain several cosmological parameters at
few percent level~\cite{Planck2}. Within the context of the
spatially flat $\Lambda$CDM cosmology, a low value of the Hubble
constant, $H_0=67.4\pm1.4~\mathrm{km}\cdot \mathrm{s}^{-1}\cdot
\mathrm{Mpc}^{-1}$, and a high value of the matter density
parameter, $\Omega_m=0.314\pm0.020$, are obtained. These are
seemingly in tension with the measurements of the magnitude-redshift
relation for Type Ia Supernova (SNe Ia)~\cite{H01, H02, SNLS3}, but
are entirely consistent with geometrical constraints from baryonic
acoustic oscillation (BAO) surveys~\cite{BAO1, BAO2}. This
inconsistency between fundamental cosmological parameters
constrained from the high redshift CMBR measurements and those from
the observations at relatively low redshifts may indicate the
existence of defects  in the cosmological model where we project
constraints on the standard cosmological parameters from these
observations to $z=0$, since projected parameters should presumably
be the same from measurements at all $z$ in a given model. Thus,
after {\it Planck}, attempts have been made to resolve this
tension~\cite{tension0, tension1, tension2,
tension3,tension4,tension5,tension6,tension7}. For instance, the
cosmic variance has been suggested to account for the discrepancy in
$H_0$~\cite{tension0} and an extension of the
Friedmann-Lem$\hat{\mathrm{a}}$itre-Robertson-Walker (FLRW) metric
to the reputed ``Swiss-cheese'' model for the background has been
proposed to alleviate the tension of $\Omega_m$~\cite{tension4}.

Here, we take a different approach to the issue. We show that if we
first calibrate, with the help of the distance-duality relation, the
light-curve fitting parameters in the distance estimation of the SNe
Ia using the data on angular diameter distance of the galaxy
clusters so as to eliminate the cosmological model-dependence that
exists in the global fit to the Hubble diagram where the light-curve
fitting parameters are treated free on the same footing as
cosmological parameters, then a higher value of the matter density
parameter $\Omega_m$ can be obtained from SNLS3. This is consistent
with the {\it Planck} at the  68.3\% confidence, thereby alleviating
the tension. Furthermore, with the light-curve fitting parameters
and the absolute magnitude of a fiducial SNe Ia calibrated first, a
low value of the Hubble constant $H_0$ which is consistent with {\it
Planck} can also be obtained.

Note that in parallel with CMBR measurements at high redshift,
accurate distance estimation to celestial objects at relatively low
redshift is another key tool in observational cosmology. Some
fundamental changes in our understanding of the universe have
resulted from such distance measurements. For example, Brahe's
supernova and Hubble's Cepheids completely reconstructed our
understanding of the cosmos~\cite{Hubblelaw}. Almost five years
after the SNe Ia were shown to be accurate standard candles,
distance measurements for them have directly led to the discovery of
the cosmic acceleration~\cite{Riess0, Perlmutter}. After several
decades of intensive study, SNe Ia remain, at present, the most
direct and mature portal to explore the essence of the accelerated
expansion~\cite{reSN}. In the past decade or so, several supernova
data sets with hundreds of well-measured SNe Ia were released, such
as ``ESSENCE"~\cite{Essence}, ``Constitution"~\cite{Constitution},
``SDSS-II"~\cite{SDSS}, and ``Union2.1"~\cite{Union21}. Since the
SNe Ia has been proposed as a distance indicator, various empirical
approaches (known as light-curve fitters) to distance estimation,
using light-curve shape parameters ($\Delta m_{15}$ or a stretch
factor)~\cite{shape1,shape2,shape3} or color
information~\cite{color1,color2}, or
both~\cite{both1,both2,both3,both4,both5}, have been advanced.
Currently, the distance of the SNe Ia is usually estimated by
expressing it as an empirical function of the observable quantities
because of the variability of the large spectra features. Taking the
SALT2 light-curve fitter~\cite{both4} as an example, the distance
estimator (distance modulus:
$\mu=5\log\big[\frac{d_\mathrm{L}}{\mathrm{Mpc}}\big]+25$) of the
SNe Ia is given by a linear combination of $m_B^*$, $x_1$, and $c$:
\begin{equation}\label{eq1}
\mu_\mathrm{B}(\alpha, \beta; M)=m_\mathrm{B}^*-M+\alpha*x_1-\beta*c
\end{equation}
where $x_1$ is the stretch (a measurement of the shape of the SNe
light curve) and $c$ is the color measurement for the SNe.
$m_\mathrm{B}^*$ is the rest-frame peak magnitude of an SNe.
$\alpha$ and $\beta$ are nuissance parameters which characterize the
stretch-luminosity and color-luminosity relationships, reflecting
the well-known broader-brighter and bluer-brighter relationships,
respectively. The value of $M$ is another nuissance parameter
representing the absolute magnitude of a fiducial SNe. In general,
in SALT2 (similar for SiFTO~\cite{both5}, or SALT2/SiFTO
combined~\cite{SNLS3}), $\alpha$ and $\beta$ are left as free
parameters (on the same weight as cosmological parameters) that are
determined in the global fit to the Hubble diagram. This treatment
results in the dependence of distance estimation on cosmological
model. Thus, cosmological implications derived from the distance
estimation of the SNe Ia with the light-curve fitting parameters
determined in the global fit to the Hubble diagram are somewhat
cosmological-model-dependent.

On the other hand, besides the luminosity distance, $d_\mathrm{L}$,
measurement for the standard candle such as SNe Ia, distance
estimation for objects with known size (that is, standard ruler)
named as angular diameter distance (ADD), $d_\mathrm{A}$, is also
often employed in astronomy. Recently, an ADD sample of 25 galaxy
clusters ($0.023\leq z\leq0.784$) has been obtained by combining the
X-ray brightness and Sunyaev-Zel'dovich temperature decrements (SZ
effect~\cite{SZeffect}) observations~\cite{galaxyc}. In addition,
the three-dimensional structure of galaxy cluster was also minutely
studied in this work and it was found that the spherical hypothesis
for geometry of cluster is generally rejected. The luminosity
distance, $d_\mathrm{L}$, and ADD, $d_\mathrm{A}$, may be measured
independently by different astronomical observations from different
celestial objects, but they relate to each other by means of the
Etherington's reciprocity relation~\cite{DD1,DD2,DD3}:
\begin{equation}\label{eq2}
\frac{d_\mathrm{L}}{d_\mathrm{A}}(1+z)^{-2}=1.
\end{equation}
This relation, sometimes referred as the distance-duality (DD)
relation, is completely general and valid for all cosmological
models based on the Riemannian geometry. That is, the validity is
dependent neither on the Einstein field equation for gravity nor on
the nature of the matter-energy content of the universe. It only
requires that the source and observer be connected by null geodesic
in a Riemannian spacetime and that the number of photons be
conserved. The fundamental DD relation has played an essential role
in modern observational cosmology, for instance,
gravitational-lensing studies~\cite{gl}, the plethora of cosmic
consequences from primary and secondary temperature anisotropies of
the CMBR observations~\cite{CMBA} and analysis from galaxy cluster
observations~\cite{gc1,gc2}.

Thus, the DD relation, the validity of which is a seemingly
reasonable assumption without new physics, along with the ADD data
of galaxy clusters, provides us a natural possibility to calibrate
the light-curve fitting parameters, $\alpha$ and $\beta$, for
distance estimation in the SNe Ia observation in a
cosmological-model-independent manner before being used to estimate
the distances of the SNe Ia for cosmological analysis. In the
following, we will demonstrate that if we use $\alpha$ and $\beta$
corrected this way to re-estimate the luminosity distances of the
SNe Ia and explore cosmological implications in the framework of the
spatially flat $\Lambda$CDM cosmology,  we can obtain a higher value
of matter density parameter, $\Omega_m=0.301^{+0.033}_{-0.031}$ (the
original SNLS3 gives $\Omega_m=0.225^{+0.040}_{-0.037}$), which is
in good agreement with that obtained from the {\it Planck}
observations. Thus, tension regarding $\Omega_m$ can be alleviated
without invoking new physics or resorting to extensions of the
standard cosmological model. Furthermore, a low value of Hubble
constant, $H_0=66.0^{+0.3}_{-0.4}~\mathrm{km}\cdot
\mathrm{s}^{-1}\cdot \mathrm{Mpc}^{-1}$, can also be obtained from
SNLS3, which is consistent with {\it Planck}.

\section{Constraints on light-curve fitting parameters and cosmological implications }
\begin{figure}[htbp]
\centering
\includegraphics[width=0.5\textwidth, height=0.45\textwidth]{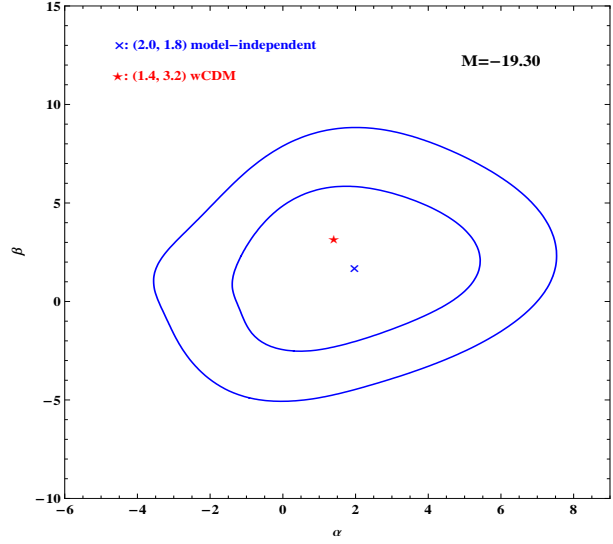}
\caption{\label{Fig1}
  Cosmological-model-independent constraint on light-curve fitting parameters, $\alpha$ and $\beta$, from SNLS3 SN Ia analyzed with
  SALT2/SiFTO combined fitter and the galaxy clusters sample.}
\end{figure}

In order to place cosmological-model-independent constraints on
$\alpha$ and $\beta$ with the aid of the reciprocity relation  in
Eq.~(\ref{eq2}), the data pairs of observed $d_\mathrm{L}$ and
$d_\mathrm{A}$ almost at the same redshift should be provided. For
the observed $d_\mathrm{L}$, the SNLS3 SN Ia sample compiled with
SALT2/SiFTO combined fitter~\cite{SNLS3} is considered. Galaxy
clusters sample, where an elliptical geometry is supposed for the
morphology of clusters and the ADDs are obtained by combining the
SZE+X-ray brightness measurements~\cite{galaxyc}, is responsible for
providing the observed $d_\mathrm{A}$. Since the sample size of the
SN Ia is much larger than that of the galaxy clusters, we bin the
observed $d_\mathrm{L}$ from the data points of the SNLS3, with the
corresponding redshifts satisfying the selecting criteria, $\Delta
z_{\mathrm{max}}=
\left|z_{\mathrm{cluster}}-z_{\mathrm{SNe~Ia}}\right|_{\mathrm{max}}\leq0.005$~\footnote{For
the galaxy cluster MS 1137.5+6625 with redshift $z=0.784$, only two
SNe Ia, 04D1jd ($z=0.778$) and 05D4cs ($z=0.79$), satisfy $\Delta
z_{\mathrm{max}}=0.006$. For the sake of completeness of the galaxy
clusters sample, these two SNe Ia are binned for matching the very
galaxy cluster.}, to match the observational data of the ADD sample,
\begin{equation}
\mu^{\mathrm{SN}}_{\mathrm{bin}}=\frac{\sum(\mu^{\mathrm{SN}}_i/\sigma_i^2)}{\sum(1/\sigma_i^2)},~~
\sigma_{\mu,\mathrm{bin}}^{\mathrm{SN}}=\bigg(\frac{1}{\sum(1/\sigma_i^2)}\bigg)^{1/2}.
\end{equation}
It should be noted that both binned distance modulus
$\mu^{\mathrm{SN}}_{\mathrm{bin}}$, and corresponding uncertainties
$\sigma_{\mu,\mathrm{bin}}^{\mathrm{SN}}$ are functions of $\alpha$
and $\beta$. In addition, we have to express the observed distances
in terms of the distance modulus, that is,
$\mu_\mathrm{B}^{\mathrm{SN}}(\alpha, \beta; M)$ for the SNe Ia
observations and
$\mu^{\mathrm{cluster}}=5\log\big[\frac{(1+z)^2d_\mathrm{A}^{\mathrm{cluster}}}{\mathrm{Mpc}}\big]+25$
for the galaxy clusters sample, to marginalize the absolute
magnitude of a fiducial SNe Ia, $M$, when $\alpha$ and $\beta$ are
fitted using the standard minimum-$\chi^2$ route,
\begin{equation}\label{chi22}
\chi^2(\alpha, \beta, M)=A-2*M*B+M^2*C\;,
\end{equation}
where
\begin{eqnarray}
A(\alpha, \beta)&=&\sum_{i=1}^{25}\frac{[\mu^{\mathrm{SN}}(z_i;
\alpha, \beta,
M=0)-\mu^{\mathrm{cluster}}(z_i)]^2}{\sigma_{\mathrm{tot}}^2(\alpha, \beta)}\;,\\
B(\alpha, \beta)&=&\sum_{i=1}^{25}\frac{[\mu^{\mathrm{SN}}(z_i;
\alpha, \beta,
M=0)-\mu^{\mathrm{cluster}}(z_i)]}{\sigma_{\mathrm{tot}}^2(\alpha, \beta)}\;,\\
C(\alpha,
\beta)&=&\sum_{i=1}^{25}\frac{1}{\sigma_{\mathrm{tot}}^2(\alpha,
\beta)}\;.
\end{eqnarray}
Here $\sigma_{\mathrm{tot}}^2$ are propagated from both the
statistical uncertainties in SNe Ia and that in galaxy clusters
observations. Eq.~(\ref{chi22}) has a minimum at $M=B/C$, and it is
\begin{equation}
\tilde{\chi}^2(\alpha, \beta)=A(\alpha, \beta)-\frac{B^2(\alpha,
\beta)}{C(\alpha, \beta)}\;.
\end{equation}
Different from the marginalization of a combination of the absolute
magnitude of a fiducial SNe Ia and the Hubble constant in the global
fit to the Hubble diagram, the analysis performed here can give an
estimation for the absolute magnitude of a fiducial SNe Ia  and thus
break the degeneracy between them.  Unfortunately, the systematic
uncertainties of the SNe Ia (in terms of the covariance matrix) are
difficult to be included when we bin the selected SNe Ia for the
corresponding galaxy cluster to obtain our data pairs. However, the
systematic errors are taken into consideration in our following
cosmological implication analysis. The
cosmological-model-independent constraint on $\alpha$ and $\beta$ is
shown in Fig.~\ref{Fig1}. Compared to the light-curve fitting
parameters determined from the global fit to the Hubble diagram in
the framework of the constant $w$ dark energy model (marked as the
red star)~\cite{SNLS3}, the result derived from our
cosmological-model-independent analysis (indicated by the blue
cross) favors a larger $\alpha$ and a smaller $\beta$. Along with
these two light-curve fitting parameters, a model-independent
estimation for the absolute magnitude of a fiducial SNe Ia
$M=-19.30$ is also achieved,  which is in good agreement with what
obtained from photometric measurements~\cite{shape1, absolute1}
(refer to review elsewhere~\cite{absolute2}). This may be seen as an
indication of reliability of our proposal to calibrate the
light-curve fitting parameters in the distance estimation for SNe Ia
using the ADD data. It is worth noting that an estimation for $M$
can not be accomplished without any assumption prior for the Hubble
constant in previous global fit procedure.
 With this estimation for the absolute magnitude of a fiducial
SNe Ia, a
constraint on Hubble constant from SNLS3 alone can be obtained in
our following analysis for cosmological implications.

\begin{figure}[htbp]
\centering
\includegraphics[width=0.75\textwidth, height=0.45\textwidth]{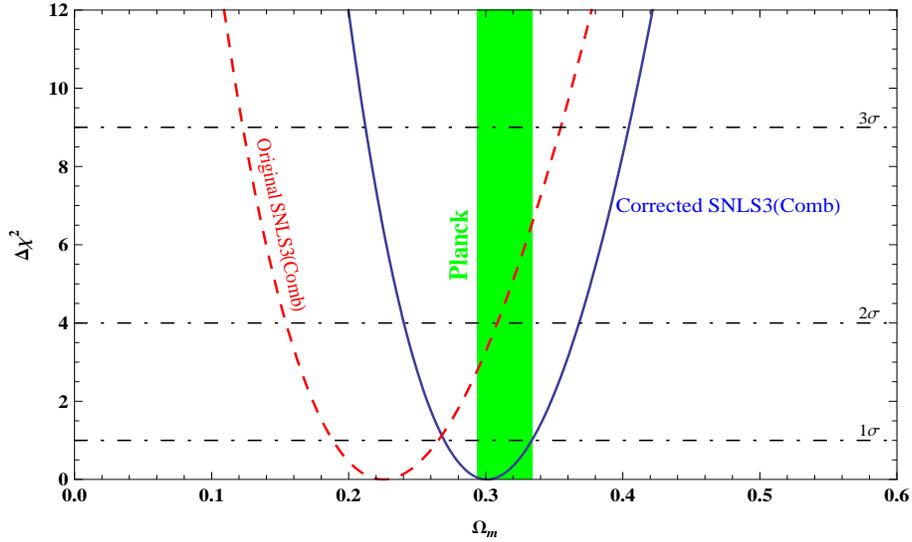}
\caption{\label{Fig2}
   Constraints on the matter density parameter, $\Omega_m$, in the context of spatially flat $\Lambda$CDM cosmology.}
\end{figure}

\begin{figure}[htbp]
\centering
\includegraphics[width=0.75\textwidth, height=0.45\textwidth]{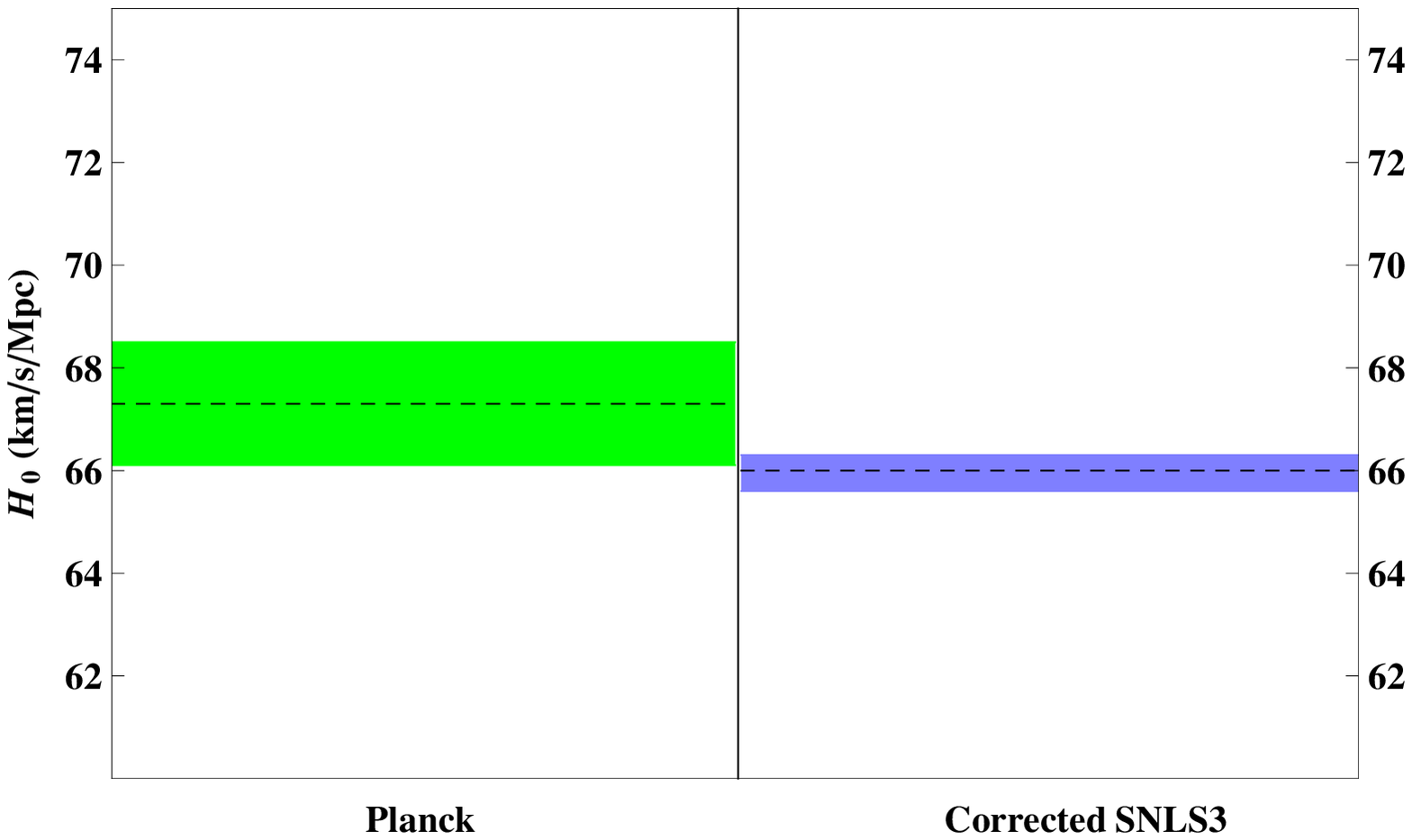}
\caption{\label{Fig3}
 Constraints on the Hubble constant, $H_0$, at 68.3\% confidence, in the context of the spatially flat $\Lambda$CDM cosmology.  }
\end{figure}

Now let us explore the cosmological implications of the corrected
distance for the SNLS3 using the best fit values for the light-curve
fitting parameters $\alpha$ and $\beta$ constrained from our
model-independent analysis. Following the minimum-$\chi^2$ route
presented in Appendix C of Ref.~\cite{SNLS3}, we place constraints
from the corrected SNLS3 SN Ia on the spatially flat $\Lambda$CDM
cosmology. The results are shown in Fig.~(\ref{Fig2}, \ref{Fig3}).
From Fig.~\ref{Fig2}, we find that, compared to the original SNLS3
SN Ia, the corrected one yields a higher value of the matter density
parameter, $\Omega_m=0.301^{+0.033}_{-0.031}$. This agrees with that
obtained from {\it Planck} observations very well. Moreover, with
the previously determined $M=-19.30$, we can also derive a
constraint on the Hubble constant from the corrected SNLS3. As shown
in Fig.~\ref{Fig3}, we obtain a low value of $H_0$
($66.0^{+0.3}_{-0.4}~\mathrm{km}\cdot \mathrm{s}^{-1}\cdot
\mathrm{Mpc}^{-1}$) which is also in agreement with that from {\it
Planck} observations at 68.3\% confidence.

\section{Conclusion And Discussion}

We have demonstrated that the tension between {\it Planck}
measurements and the observed magnitude-redshift relation for the
SNe Ia may be alleviated if we first calibrate, with help of the
distance-duality relation, the light-curve fitting parameters
$\alpha$ and $\beta$ in the distance estimation of the SNe Ia using
the data on angular diameter distance of the galaxy clusters. This
eliminates the cosmological model-dependence that exists in the
global fit to the Hubble diagram where the parameters are treated
free on the same footing as cosmological parameters. We can use
$\alpha$ and $\beta$ corrected in this manner to re-estimate the
luminosity distances of the SNLS3 SNe Ia and explore their
cosmological implications in the framework of the spatially flat
$\Lambda$CDM cosmology, and so a higher value of the matter density
parameter, $\Omega_m=0.301^{+0.033}_{-0.031}$, can be obtained. This
alleviates the tension between {\it Planck} and SNe Ia observations
regarding $\Omega_m$ significantly (rendering them be consistent at
68.3\% confidence).

Another unusual feature in our approach is that the estimation for
the absolute magnitude of a fiducial SNe Ia, $M=-19.30$, can
simultaneously be obtained from the standard minimum-$\chi^2$
fitting route for $\alpha$ and $\beta$ without any assumption for
the Hubble constant prior. This makes constraining the Hubble
constant with SNLS3 alone possible and a low value of $H_0$
($66.0^{+0.3}_{-0.4}~\mathrm{km}\cdot \mathrm{s}^{-1}\cdot
\mathrm{Mpc}^{-1}$) is obtained, which is also in good agreement
with what obtained from {\it Planck} observations. However, the
globally averaged Hubble constant we obtained here from SNLS3,
although consistent with {\it Planck}, is in tension with the
locally measured expansion rate of the universe. This might be a
result of  the cosmic variance~\cite{tension0}, or even more
speculatively, a dilute local environment~\cite{bubble1, bubble2,
bubble3}.

Finally, we must note that although the method for determining the
light-curve fitting parameters proposed here can remove the
cosmological model-dependence that is present in the global fit to
the Hubble diagram, yielding a reasonable absolute magnitude of a
fiducial SNe Ia, and thereby reducing the tensions between {\it
Planck} measurements at high red-shift and the observed
magnitude-redshift relation for the SNe Ia at relatively low
red-shifts, the presence of systematic uncertainties in measurements
using SZE+X-ray surface brightness observations and the limited
samples of the galaxy clusters may lead to biases of our result.
Thus, ADD data of more samples of galaxy clusters with greater
precision are needed to increase the statistical power of our
result. This being considered, the analysis herein presented
suggests a possibility to reconcile the {\it Planck} and SNe Ia
observations without invoking new physics or resorting to extension
of the standard cosmological model thus giving a direction for
future observational endeavors.

\section*{Acknowledgments}
We would like to thank A. Conley and A. Avgoustidis for helpful
discussions. HY thanks the Kavli Institute for Theoretical Physics
China where part of this was done for hospitality and support.  This
work was supported by the Ministry of Science and Technology
National Basic Science Program (Project 973) under Grant
No.2012CB821804, the National Natural Science Foundation of China
under Grants Nos. 10935013, 11075083, 11175093, 11222545 and
11375092, Zhejiang Provincial Natural Science Foundation of China
under Grants Nos. Z6100077 and R6110518, the FANEDD under Grant No.
200922, the Hunan Provincial Natural Science Foundation of China
under Grant No. 11JJ7001, and the SRFDP under Grant
No.20124306110001. ZL was partially supported by China Postdoc Grant
No.2013M530541.

\end{document}